\documentclass[sigconf]{acmart}

\AtBeginDocument{%
  \providecommand\BibTeX{{%
    \normalfont B\kern-0.5em{\scshape i\kern-0.25em b}\kern-0.8em\TeX}}}

\copyrightyear{2020}
\acmYear{2020}
\setcopyright{acmcopyright}\acmConference[IDEAS 2020]{24th International Database Engineering & Applications Symposium}{August 12--14, 2020}{Seoul, Republic of Korea}
\acmBooktitle{24th International Database Engineering \& Applications Symposium (IDEAS 2020), August 12--14, 2020, Seoul, Republic of Korea}
\acmPrice{15.00}
\acmDOI{10.1145/3410566.3410595}
\acmISBN{978-1-4503-7503-0/20/06}

\usepackage[ruled,linesnumbered]{algorithm2e}
\usepackage{array,multirow}
\usepackage{pifont}

\begin{document}

\title{Emotion Cognizance Improves Health Fake News Identification}

\author{Anoop K}
\affiliation{%
  \institution{University of Calicut}
  \city{Kerala}
  \country{India}}
\email{anoopk\_dcs@uoc.ac.in}

\author{Deepak P}
\affiliation{%
  \institution{Queen's University Belfast}
  \city{Northern Ireland}
  \country{UK}}
\email{deepaksp@acm.org}

\author{Lajish V L}
\affiliation{%
 \institution{University of Calicut}
 \city{Kerala}
 \country{India}}
\email{lajish@uoc.ac.in}

\renewcommand{\shortauthors}{Anoop, et al.}

\begin{abstract}
  Identifying fake news is increasingly being recognized as an important computational task with high potential social impact. Misinformation is routinely injected into almost every domain of news including politics, health, science, business, etc., among which, the fake news in the health domain poses serious risk and harm to health and well-being in modern societies. In this paper, we consider the utility of the affective character of news articles for fake news identification in the health domain and present evidence that emotion cognizant representations are significantly more suited for the task. We outline a simple technique that works by leveraging emotion intensity lexicons to develop emotion-amplified text representations and evaluate the utility of such a representation for identifying fake news relating to health in various supervised and unsupervised scenarios. The consistent and notable empirical gains that we observe over a range of technique types and parameter settings establish the utility of the emotional information in news articles, an often overlooked aspect, for the task of misinformation identification in the health domain.
\end{abstract}



\maketitle

\section{Introduction}
\label{intro}

The spread of fake news is increasingly being recognized as an enormous problem. In recent times, fake news has been reported to have grave consequences such as causing accidents \cite{ma2016detecting}, while fake news around election times has reportedly reached millions of people \cite{allcott2017social} causing concerns whether they might have influenced the electoral outcome. {\it Post-Truth} was recognized as the Oxford Dictionary Word of the Year in 2016\footnote{https://en.oxforddictionaries.com/word-of-the-year/word-of-the-year-2016}. These have spawned an extensive interest in the data analytics community in devising techniques to detect fake news in social and online media leveraging content, temporal and structural features 
(e.g., \cite{kwon2013prominent}). A large majority of research efforts on fake news detection has focused on the political domain within microblogging environments (e.g., \cite{zhao2015enquiring,qazvinian2011rumor,gayo2013predicting,ma2016detecting,ma2017detect,zubiaga2016learning}) where structural (e.g., the user network) and temporal propagation information (e.g., re-tweets in Twitter) is available in plenty. 

Fake news within the health domain have been recognized as a problem of immense significance. As a New York Times article suggests, {\it `Fake news threatens our democracy. Fake medical news threatens our lives'}\footnote{https://www.nytimes.com/2018/12/16/opinion/statin-side-effects-cancer.html}. This paper is being finalized during the times of COVID-19 when WHO has warned global citizenry against the {\it `infodemic'}\footnote{https://www.un.org/en/un-coronavirus-communications-team/un-tackling-\%E2\%80\%98infodemic\%E2\%80\%99-misinformation-and-cybercrime-covid-19}, using the term to refer primarily to fake news around the time of the pandemic. Fake health news is markedly different from fake news in politics or event-based contexts on at least two major counts. First, they originate in online websites with limited potential for dense and vivid digital footprints unlike social media channels, and secondly, the core point is conveyed through long, nuanced textual narratives. Perhaps to aid their spread, the core misinformation is often intertwined with trustworthy information. They may also be observed to make use of an abundance of anecdotes, conceivably to appeal to the readers' own experiences or self-conscious emotions (defined in \cite{tracy2004putting}). This makes health fake news detection a challenge more relevant to NLP than other fields of data analytics. In fact, techniques that totally discard content information (e.g.,~\cite{ma2017detect,wu2018tracing}) have met with reasonable success in other domains. Further, a number of fake news sub-categories such as satire, parody, and propaganda are understood to be of much less importance in health fake news (see~\cite{waszak2018spread}), making health fake news detection quite a different pursuit at the task level. 

\begin{table*}[!t]
\caption{Examples of health fake news headlines and excerpts from them}
\label{tab:examples}      
\centering
\begin{tabular}{p{15.5cm}}
\toprule
\textbf{Wi-Fi: A Silent Killer That Kills Us Slowly!} \\
WiFi is the name of a popular wireless networking technology that uses radio waves to provide wireless high-speed Internet and network connections. People can browse the vast area of internet through this wireless device.  A common misconception is that the term Wi-Fi is short for ``wireless fidelity'', however this is not the case. WiFi is simply a trademarked phrase that means IEEE 802.11x. The first thing people should examine is the way a device is connected to the router without cables. Well, wireless devices like cell phones, tablets, and laptops, emit WLAN signals (electromagnetic waves) in order to connect to the router. However, the loop of these signals harms our health in a number of ways. The British Health Agency conducted a study which showed that routers endanger our health and the growth of both, people and plants.\\
\midrule
\textbf{Russian Scientist Captures Soul Leaving Body; Quantifies Chakras} \\
It uses a small electrical current that is connected to the fingertips and takes less than a millisecond to send signals from. When these electric charges are pulsed through the body, our bodies naturally respond with a kind of `electron cloud' made up of light photons.  Korotkov also used a type of Kirlian photography to show the exact moment someone's soul left their body at the time of death! He says there is a blue life force you can see leaving the body. He says the navel and the head are the first parts of us to lose their life force and the heart and groin are the last. In other cases, he's noted that the soul of people who have had violent or unexpected deaths can manifest in a state of confusion and their consciousness doesn't actually know that they have died.\\
\midrule
\textbf{Revolutionary juice that can burn stomach fat while sleeping} \\
Having excess belly fat poses a serious threat to your health. Fat around the midsection is a strong risk factor for heart disease, type 2 diabetes, and even some types of cancers. Pineapple-celery duo is an ideal choice for those wanting to shed the fat deposits around the stomach area due to the presence of enzymes that stimulate the fat burning hormones. All you need to do is drink this incredible burn-fat sleeping drink and refrain from eating too much sugar and starch foods during the day.\\
\bottomrule
\end{tabular}
\end{table*}

We target detection of health fake news within quasi conventional online media sources which contain information in the form of articles, with content generation performed by a limited set of people responsible for it. We observe that the misinformation in these sources is typically of the kind where scientific claims or content from social media are exaggerated or distilled either knowingly or maliciously (perhaps to attract eyeballs). Example headlines and excerpts from health fake news articles we crawled is shown in Table~\ref{tab:examples}. These illustrate, besides other factors, the profusion of trustworthy information within them and the abundantly emotion-oriented narrative they employ. Such sources resemble newspaper websites in that consumers are passive readers whose consumption of the content happens outside social media platforms. This makes fake news detection a challenging problem in this realm since techniques are primarily left to work with just the article content - as against within social media where structural and temporal data offer ample clues - in order to determine their veracity.

\subsection{Our Contribution}
\label{sec:1.1}
In this paper, we consider the utility of the affective character of article content for health fake news detection, a novel direction of inquiry though related to the backdrop of fake news detection approaches that target exploiting satire and stance \cite{rubin2016fake,chopra2017towards}. We posit that \textit{fake and legitimate health news articles espouse different kinds of affective character that may be effectively utilized to improve fake news detection}. We develop a method to enrich emotion information within documents by leveraging emotion lexicons, which we informally refer as `emotion amplification'. Our emotion-enrichment method is intentionally of simple design in order to empirically illustrate the generality of the point that emotion cognizance improves health fake news detection within both supervised and unsupervised settings. 

\begin{figure*}[!t]
\centering
\includegraphics[width=0.85\linewidth]{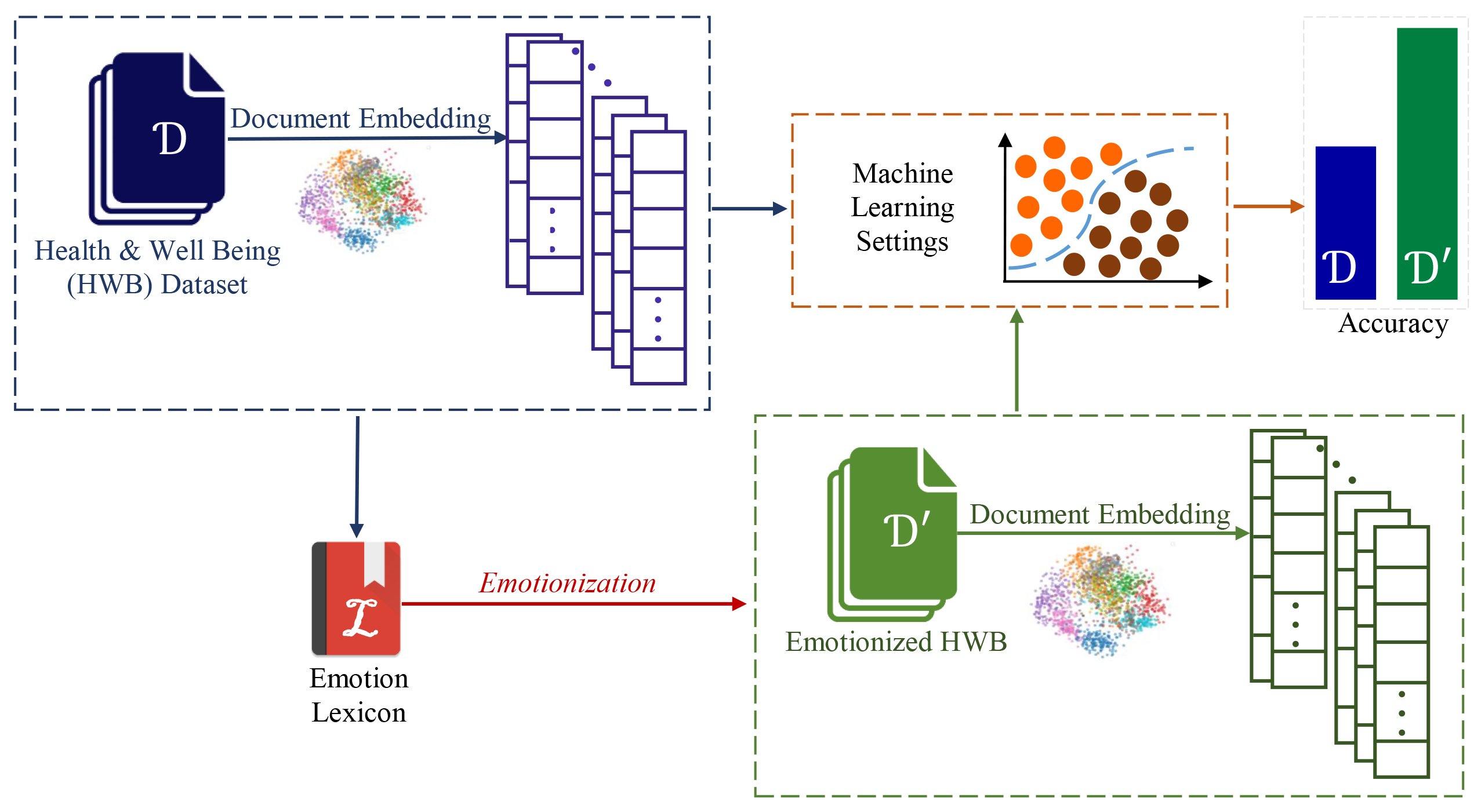}
\caption{Illustration of the empirical study}
\label{fig:arch} 
\end{figure*}

While the influence of emotions on persuasion has been discussed in recent studies \cite{vosoughi2018spread,majeed2017want}, our work provides the first focused data-driven analysis and quantification of the relationship between emotions and health fake news. Through illustrating that there are significant differences in the emotional character of fake and legitimate news in the health domain in that exaggerating the emotional content aids techniques that would differentiate them, our work sets the stage for further inquiry into identifying the nature of the differences in the emotional content. In short, we devise a methodology to leverage external emotion lexicons to derive emotion-enriched textual documents. Our empirical evaluation depicted in Figure~\ref{fig:arch} using these emotion-enriched documents for supervised and unsupervised fake news identification tasks establish that emotion cognizance improves the accuracy of fake news identification. This study is orthogonal but complementary to efforts that rely heavily on non-content features (e.g., \cite{wu2018tracing}).

\section{Related Work}
\label{sec:rel}
Our particular task, that of understanding the prevalence of emotions and its utility in detecting fake news in the health domain, has not been subject to much attention from the scholarly community. Herein, we survey two streams of related work very pertinent to our task, that of general fake news detection, and secondly, those relating to the analysis of emotions in fake news. 

Owing to the emergence of much recent interest in the task of fake news detection, there have been many publications on this topic in the last few years. A representative and non-comprehensive snapshot of work in the area appears in Table~\ref{tab:lit}. As may be seen therein, most efforts have focused on detecting misinformation within microblogging platforms using the content, network (e.g., user network) and temporal (e.g., re-tweets in Twitter) features in supervised and unsupervised settings \cite{anoop2019leveraging,wu2018tracing,ma2017detect,zhang2017detecting,zhang2016distance}; some of them, notably \cite{wu2018tracing}, target scenarios where the candidate article itself resides outside the microblogging platform, but classification task is largely dependent on information within. An emerging trend, as exemplified by \cite{wu2018tracing,ma2017detect}, focuses on how information propagates within the microblogging platform, to distinguish between misinformation and legitimate ones. Unsupervised misinformation detection techniques \cite{zhang2017detecting,zhang2016distance} start with the premise that misinformation is rare and of differing character from the large majority, and use techniques that resemble outlier detection methods in flavor.

\begin{table*}[!t]
\caption{Overview of Related Literature}
\label{tab:lit}      
\centering
    \begin{tabular}{llcccc}
    \toprule
{Work} & {Task Setting} & {Target Domain} & \multicolumn{3} {c} {{Features Used}} \\
\cline{4-6}
    &  &  & {Content} & {Network} & {Temporal}   \\
\midrule 
    Kwon et al., \cite{kwon2013prominent} & Supervised & Twitter & \ding{51} & \ding{51} & \ding{51} \\
    Zubiaga et al., \cite{zubiaga2017exploiting} & Supervised & Twitter & \ding{51} & \ding{51} & \ding{51} \\
    Qazvinian et al., \cite{qazvinian2011rumor} & Supervised & Twitter & \ding{51} & \ding{51} & \ding{51} \\
    Wu and Liu, \cite{wu2018tracing} & Supervised & Twitter & \ding{51} & \ding{51} & \ding{51} \\
    Ma et al., \cite{ma2016detecting} & Supervised & Twitter & \ding{51} & \ding{55} & \ding{51} \\
    Zhao et al., \cite{zhao2015enquiring} & Supervised & Twitter & \ding{51} & \ding{55} & \ding{51} \\
    Ma et al., \cite{ma2017detect} & Supervised & Twitter & \ding{51} & \ding{55} & \ding{51} \\
    Guo et al., \cite{guo2019exploiting} & Supervised & Weibo & \ding{51} & \ding{51} & \ding{51} \\
    Zhang et al., \cite{zhang2017detecting} & Unsupervised & Weibo & \ding{51} & \ding{55} & \ding{51} \\
    Zhang et al., \cite{zhang2016distance} & Unsupervised & Weibo & \ding{51} & \ding{55} & \ding{51} \\
    \bottomrule
    \end{tabular}
\end{table*}

\subsection{Fake News Detection}
\label{sec:2.1}

Of particular interest are recent works, \cite{bhutani2019fake} that make use of sentiment scores and \cite{guo2019exploiting} that targets to exploit emotions for fake news detection within microblogging platforms by extensive usage of {\it publisher emotions} (emotions expressed in the content) and {\it social emotions} (emotions expressed in the responses) to improve upon the state-of-the-art in fake news detection accuracies.
To contrast with these stream of works on fake news detection, it may be noted that our focus is on the health domain where information is usually in the form of long textual narratives, with limited information on the responses, temporal propagation, and author/spreader/reader network structure available for the technique to make a veracity decision.

\subsection{Emotions and Fake News}
\label{sec:2.2}

Fake news is generally crafted with the intent to mislead, and thus narratives powered with strong emotion content may be naturally expected within them. The work in \cite{bakir2018fake} analyze fake news vis-a-vis emotions and argue that what is most significant about the contemporary fake news furore is what it portends: the use of personally and emotionally targeted news produced by journalism referring to what they call as ``empathic media''. They further go on to suggest that the commercial and political phenomenon of empathically optimised automated fake news is on near-horizon, and is a challenge needing significant attention from the scholarly community. 

A recent study, \cite{paschen2019investigating}, conducts an empirical analysis on 150 real and 150 fake news articles from the political domain and report finding significantly more negative emotions in the titles of the latter. Apart from being distinctly different in terms of domain, our focus being health (vs. politics for them), we also significantly differ from them in the intent of research; our work is focused not on identifying the tell-tale emotional signatures of real vis-a-vis fake news, but on providing empirical evidence that there are differences in emotional content which may be exploited through simple mechanisms such as word-addition-based text transformations. In particular, our focus is on establishing that there are differences, and we keep the identification of the nature of differences outside the scope of our present investigation. 

A recent tutorial survey on fake news in social media, \cite{shu2019detecting}, places significant emphasis on the importance of emotional information within the context of fake news detection. On a related note, there has been some recent work~\cite{patro2019characterizing} on using emotional cues within tweets that report exaggerated health news content; it may be noted that the emotion analysis, in this case, is performed on the tweets and not on the articles themselves, making it markedly different from our work in scope.

\subsection{Our Work in Context} 
\label{sec:2.3}
To put our work in context, we note that the affective character of the content has not been a focus of health fake news detection so far, to our best knowledge. Our effort is orthogonal but complementary to most work described above in that we provide evidence that emotion cognizance in general, and our emotion-enriched data representations in particular, are likely to be of much use in supervised and unsupervised health fake news identification. As observed earlier, identifying the nature of emotional differences between fake and real news in the health domain is outside the scope of our work, but would evidently lead to interesting follow-on work.

\section{Emotionizing Text}
\label{sec:3}
The intent in this paper is to provide evidence that the affective character of fake news and legitimate articles differ in a way that such differences can be leveraged to improve the task of fake news identification. First, we outline our methodology to leverage an external emotion lexicon to build emotion amplified (i.e., {\it emotionized}) text representations. The methodology is designed to be very simple to describe and implement, so any gains out of emotionized text derived from the method can be attributed to emotion-enrichment in general and not to some nuances of the method details, as could be the case if the transformation method were to involve sophisticated steps. The empirical analysis of our emotionized representations {\it vis-a-vis} raw text for fake news identification will be detailed in the next section.

\subsection{The Task}
\label{sec:3.1}
The task of emotionizing is to leverage an emotion lexicon $\mathcal{L}$ to transform a text document $D$ to an emotionized document $D'$. We would like $D'$ also to be similar in format to $D$ in being a sequence of words so that it can be fed into any standard text processing pipeline; retaining the document format in the output, it may be noted, is critical for the uptake of the method. In short:
\begin{center}
$
D, \mathcal{L} \xrightarrow[]{Emotionization} D'
$
\end{center}
Without loss of generality, we expect that the emotion lexicon $\mathcal{L}$ would comprise of many 3-tuples, e.g., $[w, e, s]$, each of which indicate the affinity of a word $w$ to an emotion $e$, along with the intensity quantified as a score $s \in [0,1]$. An example entry could be $[unlucky, sadness, 0.7]$ indicating that the word {\it unlucky} is associated with the {\it sadness} emotion with an intensity of 0.7. 

\subsection{Methodology}
\label{sec:3.2}
Inspired by recent methods leveraging lexical neighborhoods to derive word \cite{mikolov2013distributed} and document \cite{le2014distributed} embeddings, we design our emotionization methodology as one that alters the neighborhood of highly emotional words in $D$ by adding emotion labels. As illustrated in Algorithm~\ref{algorithm}, we sift through each word in $D$ in order, outputting that word followed by its associated emotion from the lexicon $\mathcal{L}$ into $D'$, as long as the word emotion association in the lexicon is stronger than a pre-defined threshold $\tau$. In cases where the word is not associated with any emotion with a score greater than $\tau$, no emotion label is output into $D'$. In summary, $D'$ is an `enlarged' version of $D$ where every word in $D$ that is strongly associated with an emotion additionally being followed by the emotion label. This ingestion of `artificial' words is similar in spirit to {\it sprinkling} topic labels to enhance text classification \cite{hingmire2014sprinkling}, where appending topic labels to document is the focus. Table~\ref{tab:emotionize_example} shows the emotionized version of the sample article excerpts given in Table~\ref{tab:examples}.

\begin{algorithm}[!b]
\small
\SetKwInOut{Input}{input}
\SetKwInOut{Output}{output}
\SetKwInOut{Initialization}{Initialize}
\Input{Document $D$, Emotion-Lexicon $\mathcal{L}$, Parameter $\tau$}
\Output{Emotionized Document $D'$}
\BlankLine
\caption{Emotionization \label{algorithm}}
Let $D = [w_1, w_2, \ldots, w_n]$ \;
initialize $D'$ to be empty \;
\For{$(i = 1;\ i \leq n;\ i++)$}{
    write $w_i$ as the next word in $D'$ \;
    \uIf{$(\exists [w_i,e,s] \in \mathcal{L} \wedge s \geq \tau)$}{
    write $e$ as the next word in $D'$ \;
}
}
output $D'$
\end{algorithm}
\noindent

\begin{table*}[!t]
\caption{Emotionized Health Fake News Excerpts (added emotion labels in bold)}
\label{tab:emotionize_example}       
\centering
\begin{tabular}{p{15.5cm}}
\toprule
Wi-Fi: A Silent Killer {\bf fear} That Kills {\bf fear} Us Slowly! \\
WiFi is the name of a popular wireless networking technology that uses radio waves to provide wireless high-speed Internet and network connections. People can browse the vast area of internet through this wireless device.  A common misconception {\bf fear} is that the term Wi-Fi is short for ``wireless fidelity'', however this is not the case. WiFi is simply a trademarked phrase that means IEEE 802.11x. The first thing people should examine is the way a device is connected to the router without cables. Well, wireless devices like cell phones, tablets, and laptops, emit WLAN signals (electromagnetic waves) in order to connect to the router. However, the loop of these signals harms {\bf fear} our health in a number of ways. The British Health Agency conducted a study which showed that routers endanger {\bf fear} our health and the growth {\bf joy} of both, people and plants.\\
\midrule
Russian Scientist Captures Soul Leaving {\bf sadness} Body; Quantifies Chakras \\
It uses a small electrical current that is connected to the fingertips and takes less than a millisecond to send signals from. When these electric charges are pulsed through the body, our bodies naturally respond with a kind of `electron cloud' made up of light {\bf joy} photons.  Korotkov also used a type of Kirlian photography to show the exact moment someone's soul left their body at the time of death {\bf sadness}! He says there is a blue life force you can see leaving {\bf sadness} the body. He says the navel and the head are the first parts of us to lose {\bf sadness} their life force and the heart and groin are the last. In other cases, he's noted that the soul of people who have had violent {\bf anger} or unexpected deaths {\bf sandess} can manifest in a state of confusion and their consciousness doesn't actually know that they have died {\bf sadness}.\\
\midrule
Revolutionary juice that can burn stomach fat while sleeping \\
Having excess belly fat poses a serious threat {\bf anger} to your health. Fat around the midsection is a strong risk {\bf fear} factor for heart disease {\bf fear}, type 2 diabetes, and even some types of cancers {\bf sadness}. Pineapple-celery duo is an ideal choice for those wanting to shed the fat deposits around the stomach area due to the presence of enzymes that stimulate the fat burning hormones. All you need to do is drink this incredible burn-fat sleeping drink and refrain from eating too much sugar and starch foods {\bf joy} during the day.\\
\bottomrule
\end{tabular}
\end{table*}

\section{Empirical Study}
\label{sec:4}
Given our focus on evaluating the effectiveness of emotionized text representations over raw representations, we consider a variety of unsupervised and supervised methods (in lieu of evaluating on a particular state-of-the-art method) in the interest of generality. Data-driven fake news identification, much like any analytics task, uses a corpus of documents to learn a statistical model that is intended to be able to tell apart fake news from legitimate articles. Our empirical evaluation is centered on the following observation: {\it for the same analytics model learned over different data representations, differences in effectiveness (e.g., classification or clustering accuracy) over the target task can intuitively be attributed to the data representation}. In short, if our emotionized text consistently yields better classification/clustering models over those learned over raw text, emotion cognizance and amplification may be judged to influence fake news identification positively. This empirical evaluation framework is illustrated in Figure~\ref{fig:arch}. We first describe our Health and Well Being (HWB) dataset, followed by the emotion lexicon used in this work, and then the empirical study settings and their corresponding results.

\subsection{Dataset} 
\label{sec:4.1}
With most fake news datasets being focused on microblogging websites in the political domain making them less suitable for content-focused misinformation identification tasks as warranted by the domain of health, we curated a new dataset of fake and legitimate news articles within the topic of {\it health and well being} which being made publicly available at~\url{https://dcs.uoc.ac.in/cida/resources/hwb.html}. For legitimate news, we crawled $500$ health and well-being articles from reputable sources such as CNN, NYTimes, New Indian Express and, many others, manually double-checked for truthfulness. For fake news, we crawled $500$ articles on similar topics from well-reported misinformation websites such as BeforeItsNews, Nephef, MadWorldNews, and many others. These were manually verified for misinformation presence as well. Having a good mix of data sources in both fake and real categories, it may be argued, is critical to ensure that the technique is generalizable. The detailed dataset statistics is shown in Table \ref{tab:dataset}. 

\begin{table*}[!t]
\caption{Dataset Details}
\label{tab:dataset}       
\centering
\begin{tabular}{lccccc}
\toprule
\multicolumn{1}{c}{Dataset} &
  Class &
  \begin{tabular}[c]{@{}c@{}}Total Number of \\ Documents in the Class\end{tabular} &
  \begin{tabular}[c]{@{}c@{}}Average Words\\ per Document\end{tabular} &
  \begin{tabular}[c]{@{}c@{}}Average Sentences \\ per Document\end{tabular} &
  \begin{tabular}[c]{@{}c@{}}Total Number \\ of Words\end{tabular} \\ 
  \midrule
\multirow{2}{*}{Health and Well Being (HWB)} &
  Real &  500 &  724 &  31 &  362117 \\ &
  Fake &  500 &  578 &  28 &  289477 \\ 
  \bottomrule
\end{tabular}
\end{table*}

\subsection{Emotion Lexicon}
\label{sec:4.2}
For the lexicon, we use the NRC Intensity Emotion Lexicon \cite{mohammad2017word} which has data in the 3-tuple form outlined earlier. For simplicity, we filter the lexicon to retain only one entry per word, choosing the emotion entry with which the word has the highest intensity. This filtering entails that each word in $D$ can only introduce up to one extra token in $D'$. To mention concrete statistics, out of 1923 word sense entries that satisfy the threshold $\tau = 0.6$, our filter-out-non-best heuristic filtered out 424 entries (i.e., 22\%); thus, only slightly more than one-fifth of entries were affected. This heuristic to filter out all-but-one entry per word was motivated by the need to ensure that document structures be not altered much (by the introduction of too many lexicon words), so assumptions made by the downstream data representation learning procedure such as document well-formedness are not particularly disadvantaged. The emotionization using the filtered corpus was seen to lengthen documents by an average of 2\%, a very modest increase in document size. To put it in perspective, only around one in fifty words triggered the lexicon label attachment step, on an average. Interestingly, there was only a slight difference in the lengthening of document across the classes; while fake news documents were seen to be enlarged by $2.2\%$ on average, legitimate news articles recorded an average lengthening by $1.8\%$. This provides very weak, but initial evidence, that fake news has slightly more emotional content than real ones.

\subsection{Supervised Setting}
\label{sec:4.2}

\subsubsection{Conventional Classifiers}
\label{sec:4.2.1}
Let $\mathcal{D} = \{ \ldots, D, \ldots \}$ be the corpus of all news articles, and $\mathcal{D}' = \{ \ldots, D', \ldots \}$ be the corresponding emotionized corpus. Each document is labeled as either fake or not (0/1). With word/document embeddings gaining increasing popularity, we use the DBOW doc2vec model\footnote{https://radimrehurek.com/gensim/models/doc2vec.html} to build vectors over each of the above corpora separately, yielding two datasets of vectors, correspondingly called $\mathcal{V}$ and $\mathcal{V}'$. While the document embeddings are learnt over the corpora ($\mathcal{D}$ or $\mathcal{D}'$), the output comprises one vector for each document in the corpus that the learning is performed over. The doc2vec model uses an internal parameter $d$, the dimensionality of the embedding space, i.e., the length of the vectors in $\mathcal{V}$ or $\mathcal{V}'$. 

Each of these vector datasets are separately used to train a conventional classifier using train and test splits within them. By conventional classifier, we mean a model such as Naive Bayes (NB), k-Nearest Neighbour (KNN), Support Vector Machine (SVM), Random Forests (RF), Decision Tree (DT) or AdaBoost (AB). The classification model learns to predict a class label (one of {\it fake} or {\it real}) given a $d$-dimensional embedding vector. We use multiple train/test splits for generalizability of results where the chosen dataset (either $\mathcal{V}$ or $\mathcal{V}'$) is partitioned into $k$ random splits (we use $k=10$); these lead to $k$ separate experiments with $k$ models learnt, each model learnt by excluding one of $k$ splits, and evaluated over their corresponding held-out split. The accuracies obtained by $k$ separate experiments are then simply averaged to obtain a single classification accuracy score for the chosen dataset ($Acc(\mathcal{D})$ and $Acc(\mathcal{D}')$ respectively). Accuracy, a popular measure\footnote{https://developers.google.com/machine-learning/crash-course/classification/accuracy} to evaluate classifiers in binary classification scenarios such as ours, simply measures the sum of true positives and true negatives, and expresses it as a percentage of the dataset size. The quantum of improvement achieved, i.e., $Acc(\mathcal{D}') - Acc(\mathcal{D})$ is illustrative of the improvement brought in by emotion cognizance.

\subsubsection{Neural Networks}
Neural network models such as LSTMs and CNNs are designed to work with vector sequences, one for each word in the document, rather than a single embedding for the document. This allows them to identify and leverage any existence of sequential patterns or localized patterns respectively, in order to utilize for the classification task. These models, especially LSTMs, have become very popular for building text processing pipelines, making them pertinent for a text data oriented study such as ours. 

Adapting from the experimental settings for the conventional classifiers in Section~\ref{sec:4.2.1}, we learn LSTM and CNN classifiers with learnable word embeddings where each word would have a length of either $100$ or $300$. Unlike 
conventional classifiers where the document embeddings are learnt separately and then used in a classifier, this model interleaves training of the classifier and learning of the embeddings, so the word embeddings are also trained, in the process, to benefit the task. The overall evaluation framework remains the same as before, with the classifier-embedding combo being learnt separately for $\mathcal{D}$ and $\mathcal{D}'$, and the quantum by which $Acc(\mathcal{D}')$ surpasses $Acc(\mathcal{D})$ used as an indication of the improvement brought about by the emotionization.

\subsubsection{Results and Discussion}
Table~\ref{tab:classification} lists the classification results of the conventional classifiers as well as those based on CNN and LSTM, across two values of $d$ and various values of $\tau$. $d$ is overloaded for convenience in representing results; while it indicates the dimensionality of the document vector for the conventional classifiers, it indicates the dimensionality of the word vectors for the CNN and LSTM classifiers. Classification {\it models learned over the emotionized text are seen to be consistently more effective for the task}, as exemplified by the higher values achieved by $Acc(\mathcal{D}')$ over $Acc(\mathcal{D})$ (highest values in each row are indicated in bold). While gains are observed across a wide spectrum of values of $\tau$, the gains are seen to peak around $\tau \approx 0.6$. Lower values of $\tau$ allow words of low emotion intensity to influence $D'$ while setting it to a very high value would add very few labels to $D'$ (at the extreme, using $\tau=1.0$ would mean $D=D'$). Thus the observed peakiness is along expected lines, with $\tau \approx 0.6$ achieving a middle ground between the extremes. 

The quantum of gains achieved, i.e., $|Acc(\mathcal{D}')-Acc(\mathcal{D})|$, is seen to be notable, sometimes even bringing $Acc(\mathcal{D}')$ very close to the upper bound of $100.0$; this establishes that emotionized text is much more suitable for supervised misinformation identification. It is further notable that the highest accuracy is achieved by AdaBoost as against the CNN and LSTM models; this may be due to the lexical distortions brought about addition of emotion labels limiting the emotionization gains in the LSTM and CNN classifiers that attempt to make use of the word sequences explicitly. The best accuracy achieved over $\mathcal{D}'$ at $\tau = 0.6$ is {\it 96.5}, which is better than the best accuracy achieved for $\mathcal{D}$ by $6$ percentage points. 

\begin{table*}[t]
\caption{Classification Results (The numbers are between 0 to 100, and could be interpreted as percentages)}
\label{tab:classification}      
\centering
\begin{tabular}{lcccccc}
\toprule
    \multirow{2}{*}{{\small Method}} & \multirow{2}{*}{$Acc(\mathcal{D})$} &  \multicolumn{5}{c}{$Acc(\mathcal{D}')$} \\ 
    \cline{3-7} 
    & & $\tau=0.0$ & $\tau=0.2$ & $\tau=0.4$ & $\tau=0.6$ & $\tau=0.8$ \\
\midrule
    \multicolumn{7}{c}{Classification Results for $d = 100$} \\
\midrule
    NB & 77.0 & 78.0 & 78.0 & 78.5 & {\bf 79.0} & 77.5 \\
    KNN & 75.0 & 75.0 & 75.5 & 76.0 & {\bf 92.5} & 75.0 \\
    SVM & 50.0 & 65.0 & 75.0 & 75.0 & {\bf 90.0} & 70.0 \\
    RF & 63.0 & 71.0 & 70.0 & 72.0 & {\bf 84.0} & 80.5 \\
    DT & 68.0 & 69.0 & 70.0 & 78.0 & {\bf 94.0} & 78.5 \\
    AB & 55.0 & 57.0 & 70.0 & 71.0 & {\bf 96.5} & 82.5 \\
    CNN & 87.0 & 88.0 & 90.0 & 88.0 & {\bf 91.0} & 88.0 \\
    LSTM & 90.5 & 90.0 & 91.0 & 91.0 & {\bf 92.0} & {\bf 92.0} \\
\midrule
    \multicolumn{7}{c}{Classification Results for $d = 300$} \\
\midrule
    NB & 77.0 & 80.0 & 81.0 & 79.0 & {\bf 83.0} & 78.0 \\
    KNN & 72.0 & 74.0 & 75.0 & 76.0 & {\bf 91.0} & 74.5 \\
    SVM & 60.0 & 67.0 & 72.0 & 74.0 & {\bf 89.0} & 72.0 \\
    RF & 65.0 & 70.0 & 73.0 & 71.5 & {\bf 82.0} & 75.0 \\
    DT & 60.0 & 65.0 & 73.0 & 78.0 & {\bf 90.5} & 75.0 \\
    AB & 55.0 & 55.0 & 72.0 & 81.0 & {\bf 94.5} & 75.0 \\
    CNN & 91.2 & 91.0 & {\bf 92.7} & 92.0 & 92.0 & 91.0 \\
    LSTM & 90.0 & 90.2 & 90.0 & 90.2 & {\bf 90.7} & 90.0 \\
\bottomrule
\end{tabular}
\end{table*}

\subsection{Unsupervised Setting}
\label{sec:4.3}
The corresponding evaluation for the unsupervised setting involves clustering both $\mathcal{V}$ and $\mathcal{V}'$ (Ref. Sec.~\ref{sec:4.2}) using the same method and profiling the clustering against the labels on the clustering purity measure\footnote{https://nlp.stanford.edu/IR-book/html/htmledition/evaluation-of-clustering-1.html}; as may be obvious, the labels are used only for evaluating the clustering, clustering being an unsupervised learning method. We used K-Means \cite{macqueen1967some} and DBSCAN \cite{ester1996density} clustering methods, two very popular clustering methods that come from distinct families. K-Means uses a top-down approach to discover clusters, estimating cluster centroids and memberships at the dataset level, followed by iteratively refining them. DBSCAN, on the other hand, uses a more bottom-up approach, forming clusters and enlarging them by adding proximal data points progressively. Another aspect of difference is that K-Means allows the user to specify the number of clusters desired in the output, whereas DBSCAN has a substantively different mechanism, based on neighborhood density. 

For K-Means we measured purities, averaged over 1000 random initializations, across varying values of $k$ (desired number of output clusters); it may be noted that purity is expected to increase with $k$ with finer clustering granularities leading to better purities (at the extreme, each document in its own cluster would yield a purity of $100.0$). For DBSCAN we measured purities across varying values of $ms$ (minimum samples to form a cluster); the {\it ms} parameter is the handle available to the user within the DBSCAN framework to indirectly control the granularity of the clustering (i.e., the number of clusters in the output). Analogous to the $Acc(.)$ measurements in classification, the quantum of purity improvements achieved by the emotionized text, i.e., $Pur(\mathcal{D}')-Pur(\mathcal{D})$, indicate any improved effectiveness of emotionized representations. 

We would like to note here that while there are only two labels ({\it fake} and {\it real}) that we evaluate clusters against, clusterings which comprise much more than two clusters in the output provide useful evaluation settings. This is because {\it fake} and {\it real} articles may appear as various sub-structures in the dataset; these may be intermingled, making it intuitively hard to achieve good accuracies at $k = 2$. In such scenarios where the plurality of underlying clustering structures are expected to map to a small set of labels, a human-in-the-loop process may be naturally envisaged. In this, the human would look at typical documents in each cluster, and assign it one of two labels, and in cases of ambiguous clusters, subject each document in the cluster individually to manual perusal to ascertain the label to be applied. These post-clustering pipelines are significantly advantaged if the clusters are pure (either mostly fake or mostly real), so that manual perusal of individual documents can be avoided. This makes the purity of clusterings that produce much more than two clusters a pertinent measure of interest. Even when there are only two output clusters, manual cluster appraisal and assignment of {\it fake} and {\it real} labels is unavoidable since clustering algorithms do not produce labels on their own, being unsupervised methods.

\subsubsection{Results and Discussion}
Table~\ref{tab:clustering} lists the clustering results in a format similar to that of the classification study. With the unsupervised setting posing a harder task, the quantum of improvements $|Pur(\mathcal{D}')-Pur(\mathcal{D})|$ achieved by emotionization is correspondingly lower. The trends from Table~\ref{tab:clustering} are consistent with the earlier observations in that emotionization has a positive effect, with gains peaking around $\tau \approx 0.6$. The best value achieved with $\mathcal{D}'$ at $\tau = 0.6$ is $88.7\%$, which is $3.4$ percentage points better than the best purity achieved over $\mathcal{D}$. We believe the cause of low accuracy in unsupervised setting is because most conventional combinations of document representation and clustering algorithm are suited to generate topically coherent clusters, and thus fare poorly on a substantially different task of fake news identification.

\begin{table*}[t]
\caption{Clustering Results (The numbers are between 0 to 100, and could be interpreted as percentages)}
\label{tab:clustering}       
\centering
\begin{tabular}{lcccccc}
\toprule
    \multirow{2}{*}{} & \multirow{2}{*}{$Pur(\mathcal{D})$} &  \multicolumn{5}{c}{$Pur(\mathcal{D}')$} \\ 
    \cline{3-7} 
    & & $\tau=0.0$ & $\tau=0.2$ & $\tau=0.4$ & $\tau=0.6$ & $\tau=0.8$ \\
\midrule
    k & \multicolumn{6}{c}{K-Means Clustering Results for $d=100$} \\
\midrule
    2 & 52.3 & 52.4 & 52.3 & 52.3 & {\bf 56.1} &  52.9 \\
    4 & 78.1 & 78.0 & 78.6 & 79.3 & {\bf 81.6} &  79.3 \\
    7 & 85.0 & 85.7 & 85.2 & 85.1 & {\bf 86.9} &  85.6 \\
    10 & 85.3 & 85.1 & 85.1 & 85.1 & {\bf 87.7} & 85.7 \\
    15 & 85.2 & 85.3 & 85.1 & 85.1 & {\bf 87.8} & 85.8 \\
    20 & 85.2 & 85.2 & 85.0 & 85.1 & {\bf 88.7} & 85.7 \\
\midrule
    k & \multicolumn{6}{c}{K-Means Clustering Results for $d=300$} \\
\midrule
    2 & 51.3 & 52.0 & 52.0 & 52.0 & {\bf 55.5} &  52.0 \\
    4 & 77.1 & 77.8 & 78.1 & 78.9 & {\bf 81.5} &  78.5 \\
    7 & 84.0 & 84.0 & 85.0 & 84.9 & {\bf 86.9} &  84.6 \\
    10 & 85.0 & 85.0 & 85.0 & 85.0 & {\bf 87.1} & 85.1 \\
    15 & 85.1 & 85.3 & 85.1 & 85.1 & {\bf 87.5} & 85.2 \\
    20 & 85.0 & 85.2 & 85.0 & 85.0 & {\bf 88.0} & 85.0 \\
\midrule
    ms & \multicolumn{6}{c}{DBSCAN Clustering Results for $d=100$} \\
\midrule
    20 & 61.0 & 62.0 & 62.0 & 62.0 & {\bf 65.0} &  61.9 \\
    40 & 62.7 & 65.5 & 64.5 & 58.1 & {\bf 66.5} &  65.0 \\
    60 & 71.6 & 72.1 & 72.0 & \textbf{72.5} & {\bf 72.5} &  \textbf{72.5} \\
    80 & 85.1 & 85.0 & 85.1 & 85.6 & {\bf 86.0} &  85.6 \\
    100 & 84.5 & 84.1 & 84.8 & 84.7 & {\bf 86.0} & 84.0 \\
\midrule
    ms & \multicolumn{6}{c}{DBSCAN Clustering Results for $d=300$} \\
\midrule
    20 & 61.0 & 61.5 & 61.0 & 61.0 & {\bf 63.5} &  62.0 \\
    40 & 63.5 & 66.3 & 66.5 & 66.9 & {\bf 67.0} &  65.5 \\
    60 & 67.5 & 70.1 & 70.5 & 71.0 & {\bf 71.5} &  70.0 \\
    80 & 78.0 & 81.0 & 81.9 & 82.0 & {\bf 82.5} &  80.8 \\
    100 & 75.5 & 80.0 & 80.0 & 80.0 & {\bf 80.5} & 80.0 \\
\bottomrule
\end{tabular}
\end{table*}

\section{Emotionization and COVID-19 Fake News}

As we finalize this work, many parts of the world are reeling under the COVID-19 pandemic\footnote{https://en.wikipedia.org/wiki/COVID-19\_pandemic}. The core research tasks leading to this work was completed much before COVID-19 erupted. Recently, the direful effects of fake news during the times of COVID-19 pandemic has been called an `infodemic' by WHO, significantly elevating the relevance of research into combating fake news in the health domain. However, no large-scale datasets of COVID-19 fake news have been made available in the public domain as yet. 

Among the COVID-19 fake news we have come across, which include fake news on revolutionary juices\footnote{https://thelogicalindian.com/fact-check/lemon-baking-soda-coronavirus-covid-19-kills-20488}, alcohol bath\footnote{https://www.deccanherald.com/national/from-alcohol-bath-to-no-cabbage-here-are-the-covid-19-fake-news-818383.html} and cow dung bath\footnote{https://timesofindia.indiatimes.com/city/dehradun/bathing-in-cow-dung-superstitions-abound-on-how-to-tackle-covid-19/articleshow/74998817.cms}, we have found significant presence of emotional content in the narratives, indicating the applicability of emotion-oriented fake news detection for identifying COVID-19 fake news. Much of these fake news provide false hope exploiting the widespread fear of the disease and even making targeting the disadvantaged across the economic, political, and socio-cultural spectra\footnote{https://www.orfonline.org/expert-speak/how-fake-news-complicating-india-war-against-covid19-66052/}. Towards illustrating the emotional content of COVID-19 fake news, we outline the emotionized version of a representative COVID-19 fake news in Table~\ref{tab:covid}. These preliminary qualitative observations indicate that emotion-oriented techniques could be a potential direction for data science research into tackling COVID-19 fake news. 


\begin{table}[!t]
\caption{An example of Emotionized COVID-19 Fake News (added emotion labels in bold)}
\label{tab:covid}       
\centering
\begin{tabular}{p{7.75cm}}
\toprule
Do not consent to nose swab testing! \\
Avoid {\bf fear} the Covid-19 test at all costs. These swabs may be (and probably are) contaminated \textbf{fear} with something dangerous \textbf{fear}, like viruses or something we don't understand. People should be just as concerned \textbf{fear} with the swab as they are about the vaccine. I was wondering why the PCR test for COVID-19 had to be so far back and it got me thinking...how far does it go? So I did some research and found these two pictures and overlapped them. The suprising \textbf{joy} evidence was shocking \textbf{fear}! The blood {\bf fear} brain barrier \textbf{anger} is exactly where the swab test has to be placed.\\
\toprule
\end{tabular}
\end{table}

\section{Conclusions and Future Work}
\label{sec:con}
In this paper, we considered the utility of the affective character of news articles for the task of fake news detection in the health domain. We illustrated that amplifying the emotions within a news story (and in a sense, uplift their importance) helps downstream algorithms, supervised and unsupervised, to identify health fake news better. In a way, our results indicate that fake and real news differ in the nature of emotional information within them, so exaggerating the emotional information within both stretch them further apart in any representation, helping to distinguish them from each other. In particular, our simple method to emotionize text using external emotion intensity lexicons were seen to yield text representations that were empirically seen to be much more suited for the task of identifying health fake news. 

In the interest of making a broader point establishing the utility of affective information for the task, we empirically evaluated the representations over a wide variety of supervised and unsupervised techniques and methods over varying parameter settings, across which consistent and noteworthy gains were observed. This firmly establishes the utility of emotion information in improving health fake news identification. 

\subsection{Future Work} 
Given that our study establishes that there is a notable difference between fake and real news in terms of emotional profiles, we are considering ways of computationally analyzing the nature of the difference in affective character. Further, we are considering developing emotion-aware end-to-end methods for supervised and unsupervised health fake news identification, by blending article emotion cues with collective behavior heuristics that have been effective for fake news identification (e.g.,~\cite{DBLP:conf/ht/Gangireddy0L020}). Secondly, we are considering the use of lexicons learned from data \cite{bandhakavi2014generating} which may be better suited for fake news identification in niche domains. Third, we are exploring the usage of the affective content of responses to social media posts.

\begin{acks}
The first author was supported by the Rajiv Gandhi National Fellowship (RGNF), University Grants Commission (UGC), India (RGNF-2014-15-SC-KER-79884). The second author was partially supported by Ministry of Human Resource Development, Government of India (MHRD) Scheme for Promotion of Academic and Research Collaboration (SPARC) (Project ID: P620). 
\end{acks}

\bibliographystyle{ACM-Reference-Format}
\bibliography{ref_ideas}

\appendix

\section{Empirical Study Settings}

We have used the Scikit-learn machine learning library for conventional classifiers and clustering and, Keras neural-network library for CNN and LSTM. In every method, we use default parameters other than some of the important hyperparameters listed below, to aid reproducibility.

\subsection{Conventional Classifiers}

\begin{itemize}
  \item NB: $GaussianNB$ (Gaussian Naive Bayes algorithm) with default parameters 
  \item KNN: $n\_neighbors = 2$
  \item SVM: $kernel = linear$
  \item RF: $max\_depth = 5$, $n\_estimators = 10$
  \item DT: $max\_depth = 5$
  \item AB: default parameters
\end{itemize}

\subsection{Neural Networks}
We have used the CNN model presented in~\citep{kim2014convolutional}, a neural method that has recorded good performance for text classification, with following hyper-parameters.
\begin{itemize}
  \item Filter sizes = 3, 4 and 5
  \item Number of filters = 100
  \item Embedding dimension, $d$ = 100/300 (Keras Embedding)
  \item Regularizer = l2(0.01)
  \item Optimiser = Adam
  \item Loss = Binary cross-entropy
  \item Activation function in the dense layer = Sigmoid
  \item Batch size = 32
  \item Epoch = 100
\end{itemize}
The LSTM model is constructed using a single LSTM layer followed by 2 Dense layers, with following hyper-parameters.
\begin{itemize}
  \item LSTM layer  = 100/300 LSTM units 
  \item Dense layer1 = 256 neurons + relu
  \item Dense layer2 = 1 neuron + sigmoid 
  \item Embedding dimension, $d$ = 100/300
  \item Optimiser = RMSprop
  \item Loss = Binary cross-entropy
  \item Batch size = 32
  \item Epoch = 100
\end{itemize}

\subsection{Unsupervised Setting}
\begin{itemize}
  \item K-Means: $max\_iter=500$
  \item DBSCAN: default parameters
\end{itemize}

\end{document}